\documentclass[12pt]{article}
\usepackage{graphicx}   
\usepackage{pdfpages}
\usepackage{times}
\newenvironment{sciabstract}{%
\begin{quote} \bf}
{\end{quote}}
\title{Universal quantized thermal conductance in graphene} 
\author
{Saurabh Kumar Srivastav,$^{1}$ Manas Ranjan Sahu,$^{1}$ K. Watanabe,$^{2}$ T. Taniguchi,$^{2}$\\Sumilan Banerjee,$^{1}$ and Anindya Das,$^{1\ast}$\\
\\
\normalsize{$^{1}$Department of Physics,Indian Institute of Science, Bangalore, 560012, India.}\\
\normalsize{$^{2}$National Institute of Material Science, 1-1 Namiki, Tsukuba 305-0044, Japan.}\\
\\
\normalsize{$^\ast$ E-mail:  anindya@iisc.ac.in}
}
\date{}

\begin{document} 
\baselineskip24pt
\maketitle 
\begin{sciabstract}
The universal quantization of thermal conductance provides information on the topological order of a state beyond electrical conductance. Such measurements have become possible only recently, and have discovered, in particular, that the value of the observed thermal conductance of the $\frac{5}{2}$ state is not consistent with either the Pfaffian or the anti-Pfaffian model, motivating several theoretical articles. The analysis of the experiments has been made complicated by the presence of counter-propagating edge channels arising from edge reconstruction, an inevitable consequence of separating the dopant layer from the GaAs quantum well. In particular, it has been found that the universal quantization requires thermalization of downstream and upstream edge channels. Here we measure the thermal conductance in hexagonal boron nitride encapsulated graphene devices of sizes much smaller than the thermal relaxation length of the edge states. We find the quantization of thermal conductance within 5\% accuracy for $\nu$ = $1, \frac{4}{3}, 2$ and $6$ plateaus and our results strongly suggest the absence of edge reconstruction for fractional quantum Hall in graphene, making it uniquely suitable for interference phenomena exploiting paths of exotic quasiparticles along the edge.
\end{sciabstract}

\section*{Introduction}
Measurement of quantization of thermal conductance at its quantum limit ($\kappa _{0}T, \kappa _{0} = \pi^{2}k^{2}_\mathrm{B}/3h$) and showing its universality irrespective of the statistics of the heat carriers has been an important quest in condensed matter physics since the quantization can reveal the exotic topological nature of the carriers, not accessible via electrical conductance measurement\cite{kane1997quantized,senthil2000quasiparticle}. Although, the thermal conductance has been measured for phonons\cite{schwab2000measurement}, photons\cite{meschke2006single} and fermions\cite{molenkamp1992peltier,chiatti2006quantum,jezouin2013quantum} but the definitive proof of universality of quantum limit of thermal conductance remained elusive for more than two decades\cite{pendry1983quantum,kane1996thermal,kane1997quantized,rego1999fractional} until being reported very recently in fractional quantum Hall (FQHE) of GaAs based two dimensional electron gas (2DEG)\cite{banerjee2017observed,banerjee2018observation}. However, due to soft confining potential the egde-state reconstruction leads to extra pairs of counter-propagating edges in the FQHE of GaAs\cite{chklovskii1992electrostatics,chamon1994sharp,inoue2014proliferation,zhang2014theoretical,sabo2017edge} and makes it complicated to interpret the exact value of the thermal conductance. In this case, for the particular experimental set up, the measured value of thermal conductance can vary from the theoretically\cite{kane1997quantized} predicted $(N_{d} - N_{u})\kappa _{0}T$ to $(N_{d} + N_{u})\kappa _{0}T$ depending on full thermal equilibriation to no thermal equilibriation of the counter propagating edges\cite{banerjee2017observed,banerjee2018observation}, where $N_{d}$ and $N_{u}$ are the number of downstream and upstream edges, respectively. Obtaining full thermal equilibriation at very low-temperature is quite challenging as the thermal relaxation length ($\sim$ 50 $\mu$m) could be much bigger than the typical device dimensions\cite{banerjee2017observed,banerjee2018observation}. Therefore, the precise measurement of universal thermal conductance requires a system having no such edge reconstruction. Here we demosntrate that graphene, a single carbon atomic layer, which offers unprecedented universal edge profile\cite{hu2011realizing,li2013evolution} due to atomically sharp confining potential, is an ideal platform to probe universal quantized thermal conductance and unambiguously reveal the topological order of FQHE. The sharp edge potential profile in graphene is easily realized using few nanometers thick insulating spacer like hexagonal boron nitride (hBN) between the graphene and the screening layer\cite{hu2011realizing}. Furthermore, quantum Hall (QH) state of graphene has higher symmetry in spin-valley space ($SU(4)$), which is tunable by electric and magnetic field, and thus exhibit a plethora of exciting phases, ranging from spontaneously symmetry-broken states \cite{pientka2017thermal,yang2006collective,sodemann2014broken,kharitonov2012phase,feldman2009broken,weitz2010broken,young2012spin,maher2013evidence} to protected topological states like quantum spin Hall state near the Dirac point\cite{young2014tunable}. More interestingly, compared to GaAs, bilayer graphene has several additional even-denominator quantum Hall fractions\cite{li2017even} like $\frac{-1}{2}, \frac{3}{2}, \frac{-5}{2}$ and $\frac{7}{2}$, which has topologically exotic ground states with possible non-abelian excitations and some of these exotic phases can be uniquely identified by thermal conductance measurement\cite{pientka2017thermal,kane1997quantized,senthil2000quasiparticle}.

 In this report we have carried out the thermal conductance  measurement in the integer as well as FQHE of graphene devices with channel length of $\sim$ 5$\mu$m by sensitive noise thermometry setup. We first establish the quantum limit of thermal conductance for integer plateaus of $\nu = 1, 2$ and $6$ in hBN encapsulated monolayer graphene devices gated by $SiO_2/Si$ back gate. We then further study the thermal conductance for fractional plateau of $\nu$ = $\frac{4}{3}$ in a hBN encapsulated graphene device gated by graphite back gate, such that the distance (hBN $\sim$ 20 nm) between the graphene and the gate is comparable to the magnetic length scale. We show that the values of thermal conductance for $\nu$ = $\frac{4}{3}$ and $2$ are same even though they have different electrical conductance. These results indeed show the universality of thermal conductance with its quantum limit as predicted by theory\cite{kane1997quantized}. More importantly, measuring exact value of universal quantum limit of thermal conductance in such a short graphene channel strongly suggest the absence of any edge reconstruction in graphene QH as the thermal relaxation length\cite{kumada2015shot} is much larger than the channel length. Thus, our work is an important step to measure half of a thermal conductance and to demonstrate the topological non-Abelian excitaton in graphene hybrids in future. 
 
 \begin{figure*}
\centerline{\includegraphics[width=1.0\textwidth]{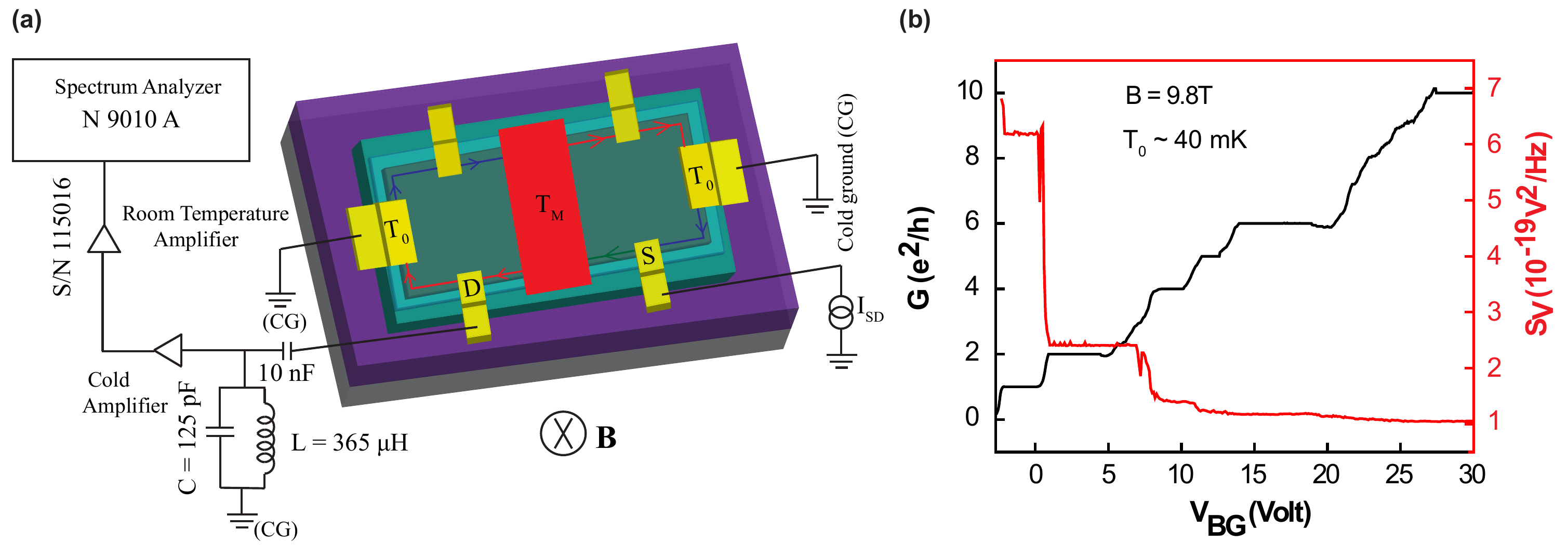}}
\caption{\textbf{Device Configuration and QH response.} (\textbf{a}) Schematic of the device with measurement setup. The device is set in integer quantum Hall regime at filling factor $\nu$=1, where one chiral edge channel (line with arrow) propagates along the edge of the sample. The current $I_{SD}$ is injected (green line) through the contact S, which is absorbed in the floating reservoir (red contact). Chiral edge channel (red line) at potential $V_{M}$ and temperature $T_{M}$ leaves the floating reservoir and terminates into two cold grounds. The cold edges (without any current) at temperature $T_{0}$  are shown by the blue lines. The resulting increase of electron temperature $T_{M}$ of the floating reservoir is determined from the measured excess thermal noise at contact D. A resonant (LC) circuit, situated at contact D, with resonance frequency $f_{0}$ = 758kHz, filter the signal, which is amplified by the cascade of amplification chain (pre-amplifier placed at 4K plate and a room temperature amplifier). Finally, the amplified signal is measured by a spectrum analyzer. (\textbf{b}) Hall conductance measured at the contact S using lock-in amplifier at B = 9.8T (black line). Thermal noise (including the cold amplifier noise) measured as function of $V_{BG}$ at $f_{0}$ = 758kHz (red line). The plateaus for $\nu$ = 1, 2 and 6 are clearly visible in both measurements.}
\label{Figure1}
\end{figure*}

 We have used two $SiO_2/Si$ and one graphite back gated devices for our measurements, where the hBN encapsulated devices are fabricated using standard dry transfer pick-up technique\cite{pizzocchero2016hot} and followed by edge contacting method (see the SI section-1). The schematic is shown in Fig.~1a, where the floating metallic reservoir in the middle connects the both sides by edge contacts. The measurements are done in a cryofree dilution refrigerator having base temperature of $\sim 12$mK. The thermal conductance was measured employing noise thermometry based on LCR resonant circuit at resonance frequency of $\sim 758$kHz and amplified by preamplifiers and finally measured by a spectrum analyzer (SI figure-2). The conductance measured at the source contact in Fig. 1a for device 1 has been plotted as a function of back gate voltage ($V_{BG}$) at B = 9.8T shown in Fig. 1b, where the clear plateaus at $\nu$=1, 2, 4, 5, 6, 10 are visible. The thermal noise (including amplifier noise) measured across the LCR circuit is plotted as a function of $V_{BG}$ in Fig. 1b, where the plateaus are also evident.

A DC current $I$, injected at the source contact (Fig. 1a), flows along the chiral edge towards the floating reservoir. The outgoing current from the floating reservoir splits into two equal parts, each propagating along the outgoing chiral edge from the floating reservoir to the cold grounds. The floating reservoir reaches a new equilibrium potential $V_{M} = \frac{I}{2 \nu G_0}$ with the filling factor $\nu$ of graphene determined by the $V_{BG}$, whereas the potential of the source contact is $V_{S} = \frac{I}{\nu G_0}$. Thus, the power input to the floating reservoir is $P_{in} = \frac{1}{2}(IV_S) = \frac{I^2}{2 \nu G_0}$, where the pre-factor of $1/2$ results due to the fact that equal power dissipates at the source and the floating reservoirs in Fig. 1a. Similarly, the outgoing power from the floating reservoir is $P_{out} = \frac{1}{2}(2 \times \frac{I}{2}V_M) = \frac{I^2}{4 \nu G_0}$. Thus, the resultant injected power dissipation in the floating reservoir due to joule heating is $J_{Q} = P_{in}-P_{out} = \frac{I^2}{4 \nu G_0}$ and as a result the electrons in the floating reservoir will get heated to a new equilibrium temperature ($T_{M}$) such that the following heat balance equation,
\begin{equation}
J_{Q}= J^{e}_{Q}(T_{M},T_{0}) + J^{e-ph}_{Q}(T_{M},T_{0}) = 0.5 N \kappa _{0}(T^{2}_{M}-T^{2}_{0}) + J^{e-ph}_{Q}(T_{M},T_{0})
\label{eqn:001}
\end{equation}
 is satisfied. Here, $J^{e}_{Q}(T_{M},T_{0})$ is the heat current carried by the $N$ chiral ballistic edge channels from the floating reservoir ($T_{M}$) to the cold ground ($T_{0}$) and the $J^{e-ph}_{Q}(T_{M},T_{0})$ is the heat loss rate from the hot electrons of the floating reservoir to the cold phonon bath. In Eq.~(\ref{eqn:001}), $T_{M}$ and $J^{e-ph}_{Q}$ are the only unknowns to determine the quantum limit of thermal conductance ($\kappa _{0}$). The $T_{M}$ of the floating reservoir in our experiment is obtained by measuring the excess thermal noise, $S_{I} = \nu k_{B}(T_{M}-T_{0})G_{0}$\cite{jezouin2013quantum,banerjee2017observed,banerjee2018observation}, along the outgoing edge channels as shown in Fig. 1a. After measuring the $T_{M}$ accurately one can determine $\kappa _{0}$ using Eq.~(\ref{eqn:001}) by tuning the number of outgoing channels ($\Delta N$).
 
 \begin{figure*}
\centerline{\includegraphics[width=1\textwidth]{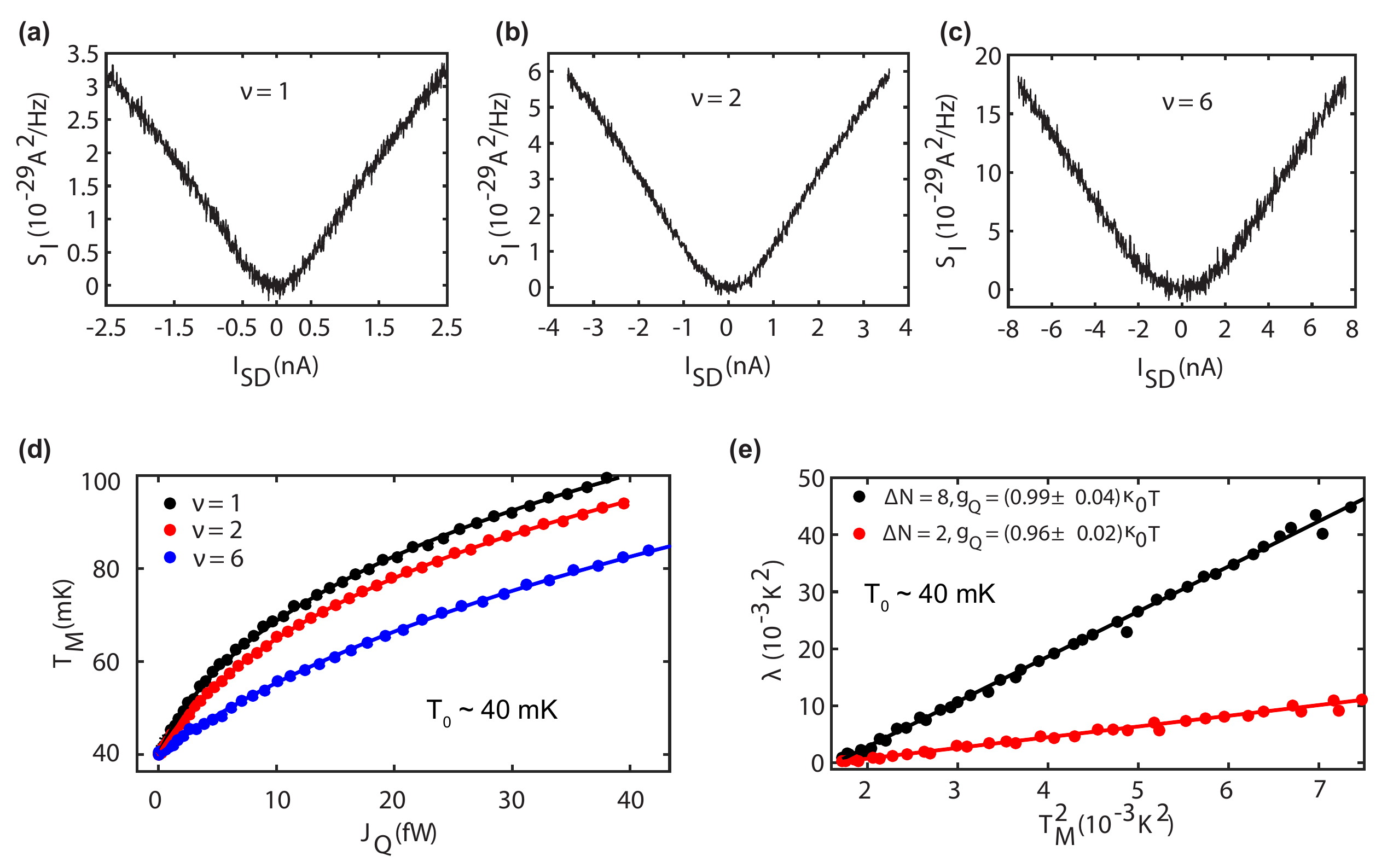}}
\caption{\textbf{Thermal conductance in integer QH.} Excess thermal noise $S_{I}$ is measured as a function of source current $I_{SD}$ at $\nu=1$ (\textbf{a}), 2 (\textbf{b}) and 6 (\textbf{c}). (\textbf{d}) The increased temperatures $T_{M}$ of the floating reservoir are plotted (solid circles) as a function of dissipated power $J_{Q}$ for $\nu$ = 1 ($N$ = 2), 2 ($N$ = 4) and 6 ($N$ = 12), respectively, where $N = 2\nu$ is the total outgoing channels from the floating reservoir in integer QH. (\textbf{e}) The $\lambda = \Delta J_{Q}/(0.5 \kappa_{0})$ is plotted as a function of $T^2_{M}$ for $\Delta N = 2$ (between $\nu$ = 1 and 2), and 8 (between $\nu$ = 2 and 6), respectively in red and black solid circles, where $\Delta J_{Q} = J_{Q}(\nu_{i},T_{M})-J_{Q}(\nu_{j},T_{M})$. The solid lines are the fittings to extract the thermal conductance values. Slope of these linear fits are 1.92 and 7.92 for $\Delta N$ = 2, and 8, respectively, which gives the $g_{Q} = 0.96 \kappa _{0}T$, and $0.99 \kappa _{0}T$ for single edge channel, respectively.}
\label{Figure2}
\end{figure*}

Fig. 2a-c shows the measured excess thermal noise $S_{I}$ for device 1 as a function of source current $I_{SD}$ for $\nu$ = 1, 2 and 6 at $B = 9.8T$. In our  experiment, for a filling factor $\nu$, the $\nu$ chiral edge modes impinge the current in the floating reservoir and $N=2\nu$ chiral edge modes leave the floating reservoir in the integer QH regime, out of them half propagate towards the left cold reservoir and other half towards the right cold reservoir. The heating of floating reservoir is exhibited in the increase of $S_{I}$ at finite current $I_{SD}$. The $x$-axis and $y$-axis of the Fig. 2 (top panel) are converted to $J_{Q}$ and $T_{M}$, respectively, and plotted in Fig. 2d for different $\nu$, where each solid circles are generated after averaging 9 consecutive data points (raw data - SI section-7). The $T_{0} \sim 40 mK$ without DC current was determined from the thermal noise measurement and shown in SI section-3. As expected we observe that $T_{M}$ is higher for lower $\nu$ as less number of chiral edges has to carry the heat away. Similarly, to maintain a constant $T_{M}$, higher $J_{Q}$ is required for higher $\nu$. In Fig. 2e we have plotted $\lambda = \Delta J_{Q}/(0.5 \kappa_{0})$ where $\Delta J_{Q} = J_{Q}(\nu_{i},T_{M})-J_{Q}(\nu_{j},T_{M})$, as a function of $T^2_{M}$ for two different configuration of $\Delta N = 2$ and $8$. It can be seen that the $\lambda$ is proportional to $T^{2}_{M}$ as expected from Eq.~(\ref{eqn:001}). The solid lines in Fig. 2e represent the linear least square fits and gives the values of 1.92 and 7.92 for $\Delta N = 2$ and $8$, respectively. Similarly, we have repeated the experiment at $B = 6T$ for device 1 and device 2 and the linear fits give values of 7.76 and 8.64 (SI figure-13 and 14) for $\Delta N = 8$, respectively. Thus, the average thermal conductance of single edge state ($\Delta N = 1$) from the four linear fittings is found to be $g_{Q} = (1 \pm 0.05)\kappa _{0}T$ upto the second place of decimal, where $T = (T_{M}+T_{0})/2$ and the error is the standard deviation.

In order to measure the thermal conductance for FQHE state we have used a graphite back gated device (device 3), where the graphene channel is isolated from the graphite gate by bottom hBN ($\sim$ 20 nm). We have also introduced extra low pass filter at the mixing chamber in order to get the lower electron temperature, $T_{0} \sim 27$mK (SI section-3). The conductance plateaus and the thermal noise as a function of $V_{BG}$ at B = 7T are shown in Fig. 3a, where the $\nu$ = 1, $\frac{4}{3}$ and 2 are clearly visible in both measurements. 
 The plots for $T_{M}$ versus $J_{Q}$ are shown in the SI figure-16. In Fig. 3b we have plotted the $J_{Q}$ (solid circles) as a function of $T^2_{M} - T^2_{0}$ for $\nu$ = 1, $\frac{4}{3}$ and 2 over the temperature window (up to $\sim$ 60-70mK) where the curve is linear, implying the dominance of the electronic contribution to the heat flow. The solid lines in Fig. 3b represent the linear fits (in $0.5\kappa_{0}$) and gives the values of 2.04, 4.16 and 4.04, which corresponds to $g_{Q} = 1.02, 2.08$ and $2.02 \kappa _{0}T$ for $\nu$ = 1, $\frac{4}{3}$ and 2, respectively. For $\nu$ = $\frac{4}{3}$, two downstream charge modes, one integer and one fractional (inner $\nu$ = $\frac{1}{3}$ with effective charge, $e^* = \frac{e}{3}$) are expected. The thermal conductance from these modes should be the same as $\nu$ = 2 having two integer downstream charge modes, as seen in our experiment within 3$\%$ mismatch. Thus, our result is consistent with the theory that the quantum limit of thermal conductance is same for both fractional and integer QH edges.
 
 \begin{figure*}
\centerline{\includegraphics[width=1\textwidth]{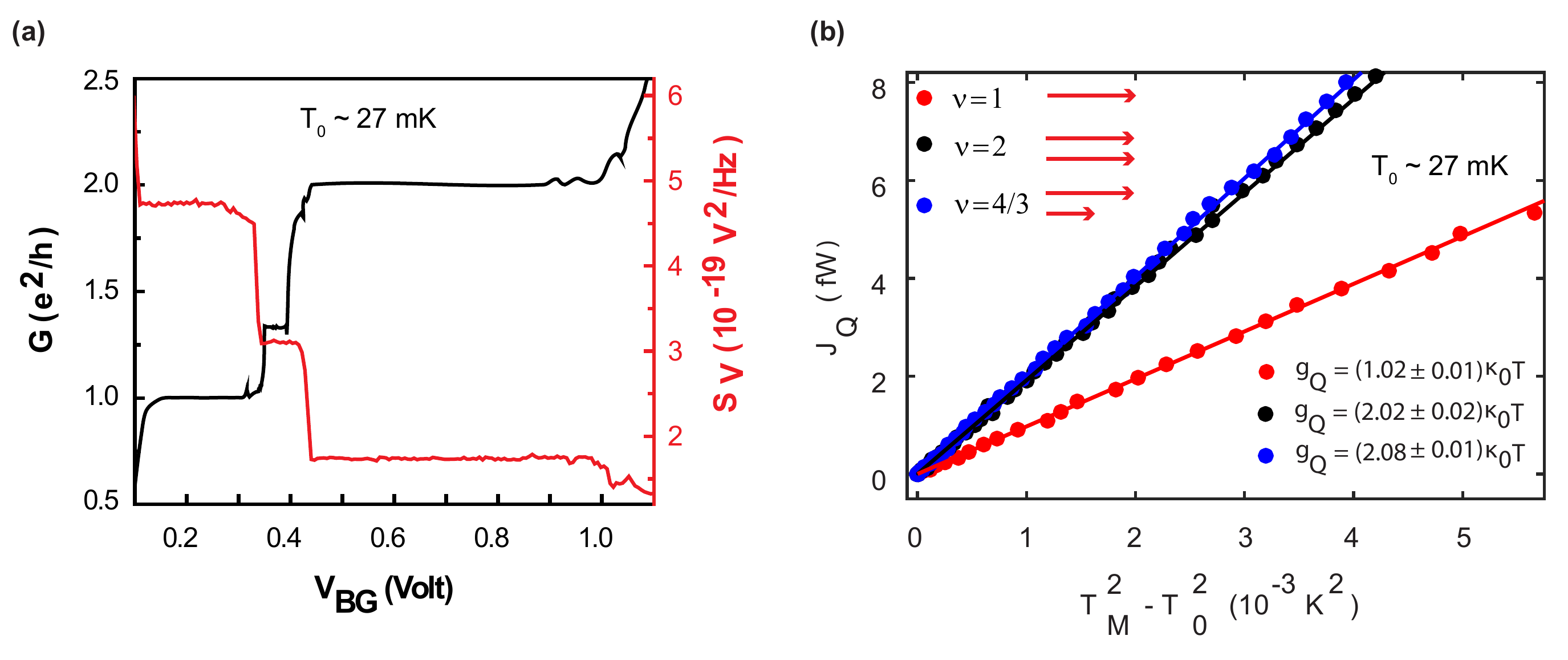}}
\caption{\textbf{Thermal conductance in fractional QH.} (\textbf{a}) Hall conductance (black line) and thermal noise (red line) measured in the graphite back gated device plotted as a function of $V_{BG}$ at B = 7T. The plateaus for $\nu$ = 1, $\frac{4}{3}$ and 2 are clearly visible in both the measurements. (\textbf{b}) Similar to the previous plots (Fig. 2), the excess thermal noise, $S_I$ is measured as a function source current $I_{SD}$ and the $T_M$ is shown as a function of the dissipated power $J_Q$ in the SI figure-15 and 16. From which we have extracted the $J_Q$ (solid circles) as a function of $T^2_{M} - T^2_{0}$ for $\nu$ = 1, 4/3 and 2 and shown upto $T_M$ $\sim$ 60-70mK. The solid lines are the linear fits to extract the slopes that give the thermal conductance values 1.02, 2.08 and 2.02$\kappa _{0}T$ for $\nu$ = 1, $\frac{4}{3}$ and 2, respectively. 
The inset shows the corresponding downstream charge modes for integer and fractional edges.} 
\label{Figure3}
\end{figure*}
 
 The reported thermal conductance in Ref\cite{banerjee2017observed,banerjee2018observation} for the particle like composite fermions ($\nu$ = $\frac{1}{3}$, $\frac{7}{3}$) matches well with the theory, 
 though its values are not consistent for the hole like composite fermions ($\nu$ = $\frac{2}{3}$, $\frac{8}{3}$) and $\nu$ = $\frac{5}{2}$. One of the main reasons behind this is the presence of extra counter-propagating edge modes arising from inevitable edge reconstruction in GaAs and insufficient thermalization among the edges. The thermal relaxation length of $\sim$ 15 $\mu$m in graphene QH is reported at 4K\cite{kumada2015shot} and it is expected to be order of magnitude higher at low-temperature ($\sim$ 50 mK). These length scales are much bigger than our graphene channel length ($\sim$ 5 $\mu$m) and thus, observing the quantum limit of thermal conductance accurately for FQHE in graphene strongly suggest the absence of edge reconstruction. In the SI section-9, we have discussed about the accuracy of our measurements and the electron-phonon coupling to the heat flow in the graphene devices.

In conclusion, we have first time measured the thermal conductance for three integer plateaus (1, 2, 6) and one particle like fractional plateau ($\frac{4}{3}$) of graphene and the values are consistent with the quantum limit ($\frac{\pi^{2}k^{2}_{B}}{3h}T$) within 5\% accuracy. These studies can be extended soon to measure the thermal conductance for the even denominator QH plateaus in graphene\cite{li2017even} with atomically sharp confining potential to probe their non-abelian nature. 


\bibliographystyle{Science}

\section*{Acknowledgments}
We would like to give special thanks to Prof. Jainendra Jain for the critical inputs as well as his help to write the manuscript. We would also like to thank Prof. Moty Heiblum for fruitful discussion. We also wish to thank Prof. Rahul Pandit, Dr. Hyungkook Choi, Dr. Yuval Ronen and Dr. Vibhor Singh for useful discussions. The authors acknowledge device fabrication and characterization facilities in CeNSE, IISc, Bangalore. A.D thanks Department of Science and Technology (DST), Government of India, under Grant nos: DSTO1470 and DSTO1597. K.W. and T.T. acknowledge support from the Elemental Strategy Initiative conducted by the MEXT, Japan and and the CREST (JPMJCR15F3), JST.

\section*{Author contributions}
S.K.S. contributed to device fabrication, data acquisition and analysis. M.R.S. contributed in noise setup, data acquisition and analysis. A.D. contributed in conceiving the idea and designing the experiment, data interpretation and analysis. S.B contributed in data interpretation and theoretical understanding of the manuscript. K.W and T.T synthesized the hBN single crystals. All the authors contributed in writing the manuscript.
\thispagestyle{empty}
\mbox{}


\section*{Supplementary material}

Supplementary Materials for

# Universal quantized thermal conductance in graphene

Saurabh Kumar Srivastav[1†], Manas Ranjan Sahu[1†], K. Watanabe[2], T. Taniguchi[2], Sumilan Banrjee[1], and Anindya Das[1*]

[1]Department of Physics,Indian Institute of Science, Bangalore, 560012, India.

[2]National Institute of Material Science, 1-1 Namiki, Tsukuba 305-0044, Japan.

*Correspondence to: anindya@iisc.ac.in

**This PDF file includes:**

# Materials and Methods

# SI section-1: Device fabrication, characterization and measurement setup

Our encapsulated graphene devices were made using the following procedures similar to those used in previous reports (*1, 2*). First, an hBN/graphene/hBN stack was made using "hot pick-up" technique (*3*). This involved the mechanical exfoliation of graphite and bulk hBN crystal on $SiO_2/Si$ wafer to obtain the single layer graphene and thin hBN ($\sim 20-30$ nm). Single layer graphene and thin hBN ($\sim 20-30$ nm) were identified using an optical microscope. Fabrication of this hetrostructures assembly involved four steps. Step1: we have used a Poly-Bisphenol-A-Carbonate (PC) coated Polydimethylsiloxane (PDMS) block mounted on a glass slide attached to tip of a micromanipulator to pick-up the exfoliated hBN flake. The exfoliated hBN flake was picked up at temperature 90°C. Step2: a previously picked-up hBN flake was aligned over a graphene. Now this graphene was picked up at temperature 90°C. Step3: bottom hBN flake was picked up using the previously picked-up hBN/Gr follwing the step2. Step4: finally, this resulting hetrostructure (hBN/Gr/hBN) was dropped down on top of an oxidized silicon wafer (p++ doped silicon with $SiO_2$ thickness of 285 nm) at temperature 140°C which served as a back gate (for graphite back gate device after the step 3, graphite flake was picked up using the previously picked-up hBN/Gr/hBN following step2. After this step, again step4 was followed). These final stacks were cleaned in chloroform ($CHCl_3$) followed by acetone and iso-propyl alcohol (IPA). The next step involved electron beam lithography (EBL) to define the contacts region. As the first step, a Poly-methyl-methacrylate (PMMA) was coated on resulting hetrostructure. Contacts region were defined on using EBL. Apart from conventional Hall probe geometry, we have defined a region for floating reservoir of $\sim$ 4 - 7 $\mu m^2$ area. We have used two $SiO_2/Si$ back gated device (device-1 and device-2) and one graphite back gated device (device-3) for the thermal conductance measurement. The edge contacts were achieved by reactive ion etching (mixture of $CHF_3$ and $O_2$ gas were used with flow rate of 40 sccm and 4 sccm, respectively at 25°C with RF power of 60W), where the etching time has been varied from 100 to 50 sec for $SiO_2/Si$ and graphite back gated devices, respectively, such that for $SiO_2/Si$ devices the bottom hBN is being etched completely where as for graphite back gated device the bottom hBN is partially etched to isolate the contacts from the bottom graphite back gate. Finally the thermal

deposition of Cr/Pd/Au (5/15/60 nm) was done to make the contacts in a evaporator chamber having base pressure of $\sim 1 - 2 \times 10^{-7}$ mbar and followed by lift-off procedure in acetone and IPA. Thus, the floating metallic reservoir in the middle are connected to the both sides graphene part by the edge contacts. This procedure of making devices prevented contamination of exposed graphene edges with polymer residues, resulting in high-quality contacts. We have measured the two terminal total resistances (R) of device-1, device-2 and device-3 (graphite back gate) as a function of gate voltage ($V_{BG}$) at zero magnetic field. The measured data is fitted with the following equation (4)

$$R = R_C + \frac{L}{We\mu\sqrt{(n_0^2 + (\frac{C_{BG}(V_{BG}-V_{CN})}{e})^2)}}$$ (1)

where $R_C$, L, W, $\mu$, and $e$ are the contact resistance, length, width, mobility, and electron charge, respectively. Carrier concentration of channel is $\frac{C_{BG}(V_{BG}-V_{CN})}{e}$ with $C_{BG}$ and $V_{CN}$ are the capacitance per unit area of back gate and voltage at the charge neutrality point, respectively. $n_0$ is the charge inhomogeneity.

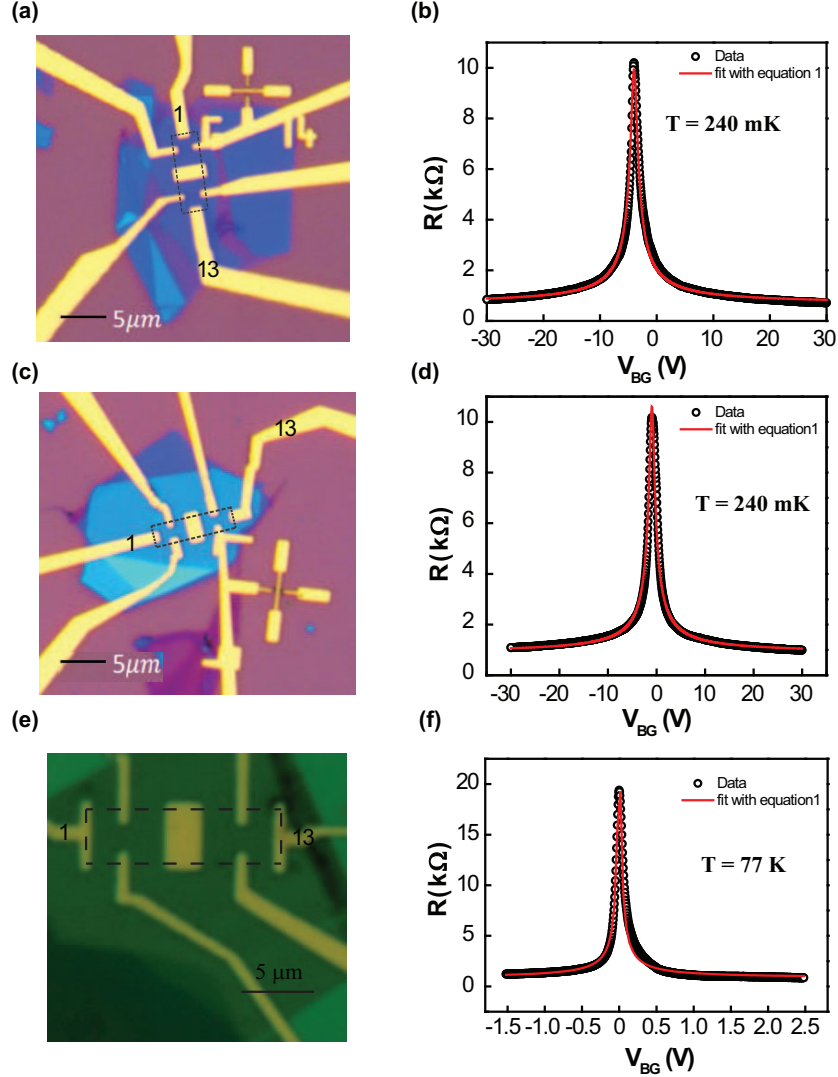

**SI Figure 1: Optical image and device response at zero magnetic field.** The one dimensional edge contacted devices were fabricated using standard dry transfer technique as discussed in the SI section-1. (**a,c**) Apart from the conventional quantum Hall (QH) geometry, there is a floating metallic reservoir of area $\sim 6\mu m^2$ and $\sim 4\mu m^2$ of device-1 and device-2, respectively. (**b,d**) Two probe gate response (measured along two extreme ends marked as 1 and 13 in fig (a,c)) at 240 mK using standard Lock-in technique. Open circle shows the data and the red curve is the fit of data in accordance to equation 1 of SI section-1. This fit gives the mobility $\sim 100000\ cm^2V^{-1}s^{-1}$ and $\sim 110000\ cm^2V^{-1}s^{-1}$ of device-1 and device-2, respectively. (**e**) The optical image of the graphite back gated device with floating reservoir area of $\sim 7\mu m^2$. (**f**) Gate response of the device-3 (graphite back gate) with mobility $\sim 100000\ cm^2V^{-1}s^{-1}$ at 77 K for hole side.

4

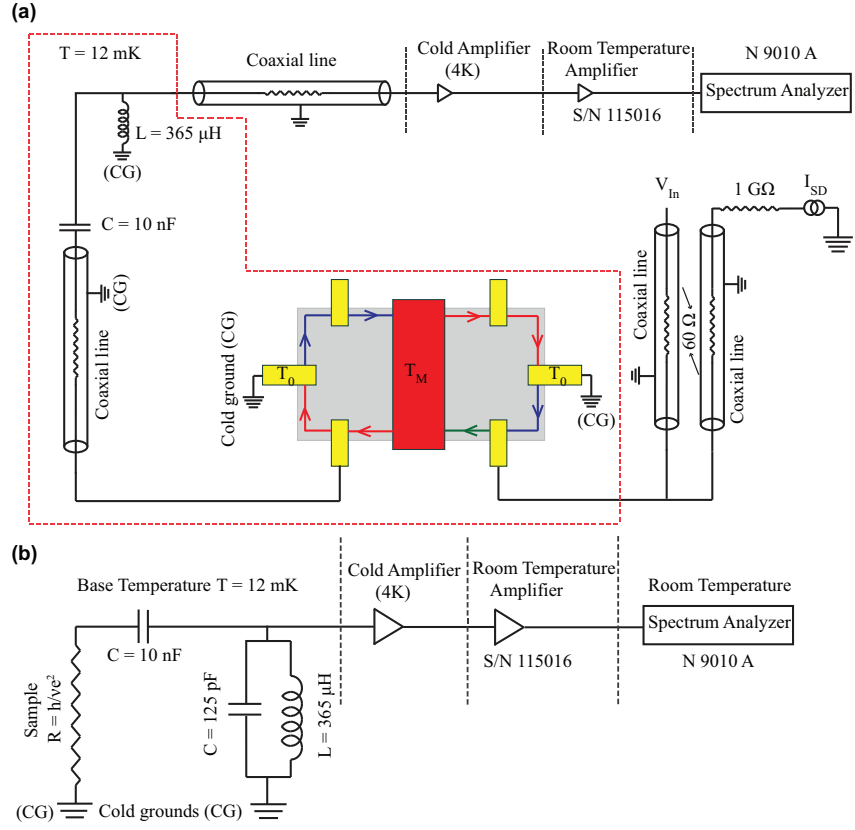

**SI Figure 2: Experimental set-up for noise measurement.** (**a**) Schematic of measurement set-up. The device was fixed to the mixing chamber plate of dilution refrigerator in Hall probe configuration with cold grounds. The sample was current biased with a 1 GΩ resistor located at the top of fridge to avoid the extra unwanted noise. Current fluctuations are converted on chip into voltage fluctuations using the well defined quantum Hall (QH) resistance $R = h/\nu e^2$ (as shown schematically in fig(**b**)) where $\nu$ is the filling factor. The signal was amplified with a home made cryogenic voltage pre-amplifier, which is thermalized to 4K plate of dilution refrigerator. This pre-amplified signal was then amplified using a voltage amplifier (PR-E3-SMA S/N 115016) placed at room temperature. After second stage of amplification, signal was measured by a spectrum analyzer (N9010A). We have used the band width of ∼ 30 kHz. Inductor L ∼ 365 $\mu$H of the L//C tank was made of an superconducting coil thermally anchored to the mixing chamber of dilution refrigerator. The parallel C ∼ 125 pF is the capacitance that develops along the coaxial lines connecting the sample to the cryogenic pre-amplifier. A ceramic capacitance of 10 nF was used between sample and inductor to block the DC current along the measurement line. The typical input voltage noise and current noise of cryogenic pre-amplifiers were ∼ 0.25-0.3 $nV/\sqrt{Hz}$ and ∼ 10-20 $fA/\sqrt{Hz}$, respectively. (**b**) The schematic circuit diagram of the measurement setup.

# SI section-2: Gain of the amplification chain

We have used two different methods to calculate the overall gain (G) of amplification chain. (i) Measurement of output voltage for a known input signal in quantum Hall states at resonance frequency; (ii) From temperature dependent Johnson-Nyquist noise (thermal noise) (*5*).

**Output voltage in quantum Hall states:** In quantum Hall (QH) regime, one can directly find the relation between source voltage and the output voltage for our device geometry. In our device geometry, the voltage probe which is the part of amplification chain will always see the equilibrium potential of floating reservoir in integr quantum Hall regime. Using Landauer-Buttiker formalism, we have derived the relation between potential of floating reservoir and source contact. In experiment, we have measured the voltage $V_{in}$ at the source contact using a seperate coaxial line as shown in SI figure 2. The output voltage $V_{out}$ is measured along the amplification chain using spectrum analyzer. To derive the relation between the output voltage $V_{out}$ and the $V_{in}$, a schematic figure (SI figure 3) is used, in which contacts '1' and '2' represent the two cold grounds. Floating contact and the source contact are shown by contacts '3' and '4', respectively. According to this nomenclature $V_{in} = V_4$. Using Landauer-Buttiker formalism (*6*), we get

$$\begin{pmatrix} I_1 \\ I_2 \\ I_3 \\ I_4 \end{pmatrix} = \begin{pmatrix} 0 & 0 & \nu G_0 & 0 \\ 0 & 0 & \nu G_0 & 0 \\ \nu G_0 & 0 & 0 & \nu G_0 \\ 0 & \nu G_0 & 0 & 0 \end{pmatrix} \begin{pmatrix} V_1 \\ V_2 \\ V_3 \\ V_4 \end{pmatrix}$$

Here $G_0$ is the quantum of conductance ( $G_0 = e^2/h$ ). From above matrix we will find following relation:

$$I_1 = \nu G_0 V_3 \tag{2}$$

$$I_2 = \nu G_0 V_3 \tag{3}$$

$$I_3 = \nu G_0 V_1 + \nu G_0 V_4 \tag{4}$$

$$I_4 = \nu G_0 V_2 \tag{5}$$

As we have used two cold grounds in our measurement setup, so in SI fig (3), 1 and 2 electrodes are grounded i.e. $V_1 = V_2 = 0$ we find,

$$I_3 = \nu G_0 V_4 \tag{6}$$

6

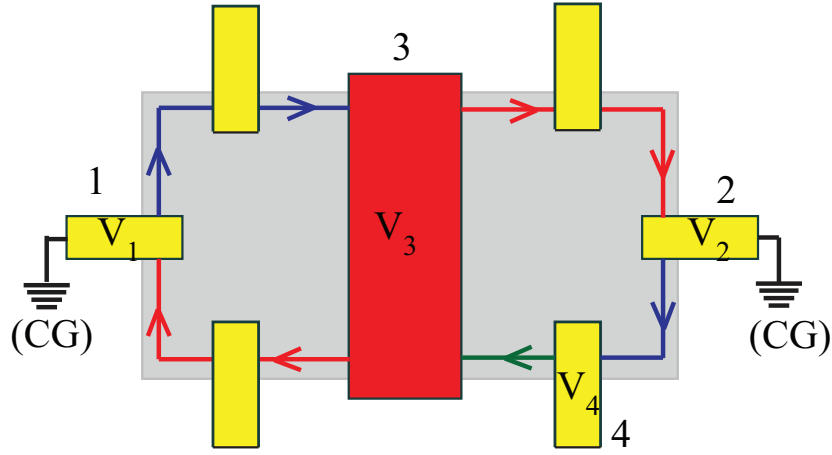

**SI Figure 3: Schematic used to derive the Gain in SI section-2**

so

$$V_4 = I_3/\nu G_0 \tag{7}$$

but $I_3$ is nothing but it is the impinging current in floating reservoir. In equilibrium,

$$I_3 = I_1 + I_2 \tag{8}$$

using equation (2), (3), (4), and (6), we obtain

$$V_3 = V_4/2 \tag{9}$$

Electrode which is connected to the amplification circuit will see the voltage $V_3$. Spectrum analyzer output will give us the voltage $V_{out}$ which is given by

$$V_{out} = GV_3 \tag{10}$$

where G is the gain of amplification chain. Now gain will be given by

$$G = V_{out}/V_3 = 2(V_{out}/V_4) = 2(V_{out}/V_{in}) \tag{11}$$

**Temperature dependent thermal noise:** In absence of any source current, equilibrium voltage noise spectrum

$$S_V = G^2(4k_BTR + V_n^2 + i_n^2R^2)BW \tag{12}$$

with $k_B$ the Boltzmann factor and T the temperature, $V_n^2$ and $i_n^2$ are the intrinsic voltage and current noise density of the amplifier, and BW is the frequency bandwidth. For a fixed integer quantum Hall plateau, change in temperature of mixing chamber (MC) plate will only affect the first term in equation (12), other terms are independent of temperature. So for two different MC plate temperature.

for temperature $T_1$

$$S_V(T_1) = G^2(4k_BT_1R + V_n^2 + i_n^2R^2)BW \tag{13}$$

and for temperature $T_2$

$$S_V(T_2) = G^2(4k_BT_2R + V_n^2 + i_n^2R^2)BW \tag{14}$$

subtracting equation (13) from (14), we get

$$G = \sqrt{\frac{S_V(T_2) - S_V(T_1)}{4k_B(T_2 - T_1)BW}} \tag{15}$$

If one plots the $\frac{S_V}{BW}$ as a function of temperature T, slopes of the linear fit will give $4G^2k_BR$, which directly gives the gain of amplification chain and from intercept, intrinsic noise of amplifier can be found.

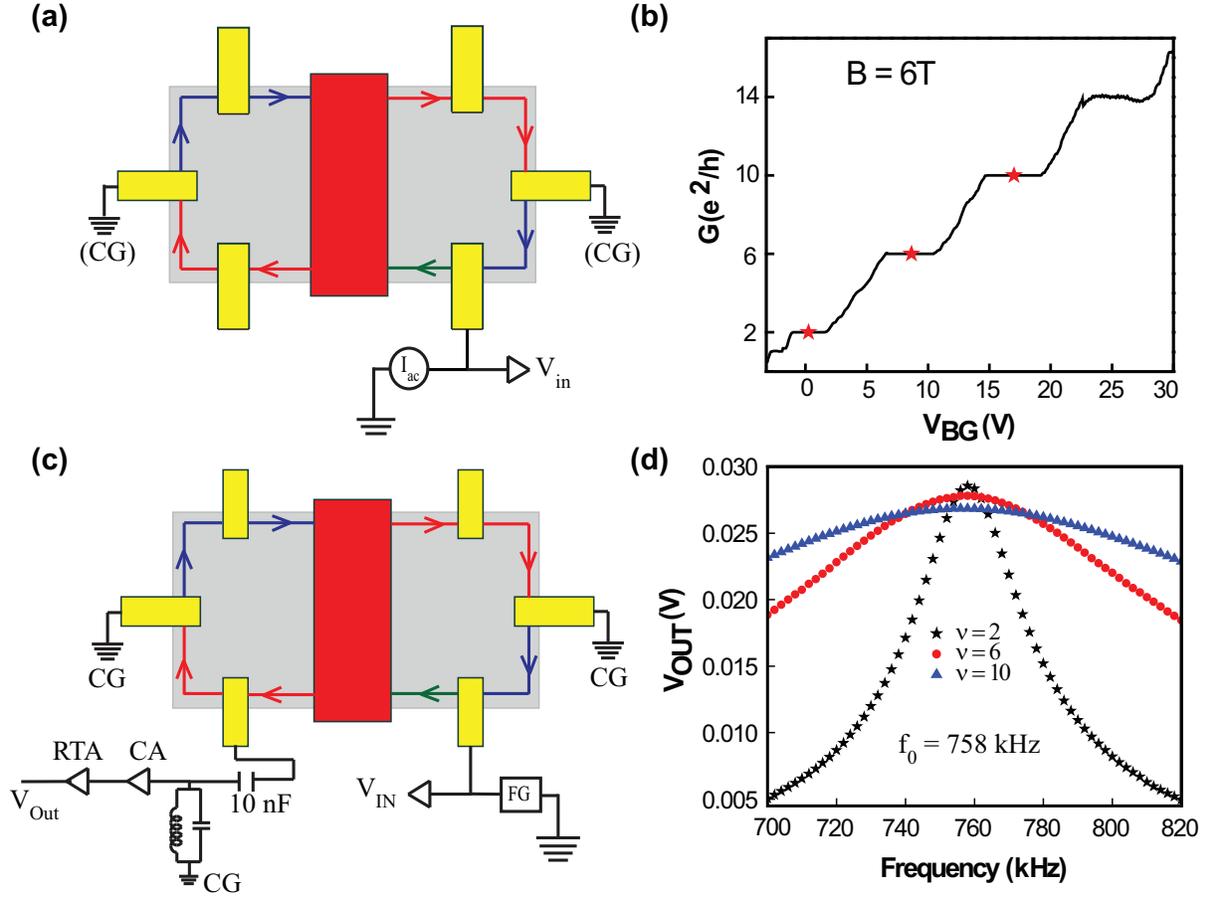

**SI Figure 4: Gain of amplification chain.** (**Output voltage from a known input signal in QH state at resonance frequency:**) (**a**) Schematic of the measurement used to measure conductance using Lock-in amplifier (228 Hz). (**b**) Gate response of Device-1 at B = 6T using low frequency Lock-in technique. Clear plateau in conductance at 2, 6 and 10 are observed in unit of $e^2/h$. Red stars display the gate voltages at which we have measured the output voltage for given input signal at high frequency. (**c**) Schematic of the measurement used to measure the output voltage for given input voltage in quantum Hall regime at high frequency (758 kHz). A high frequency signal is applied with function generator and voltage $V_{in}$ is measured using the co-axial line as shown in SI figure 2). CA and RTA stand for "cold amplifier" and "room temperature amplifier", respectively. (**d**) Output voltage $V_{Out}$ is measured using spectrum analyzer and is plotted as a function of frequency for filling factors of $\nu = 2$, 6 and 10. From the output voltage at resonance frequency, gain is calculated using $G = 2(\frac{V_{Out}}{V_{in}})$ (see SI section 2). Calculated value of gain is shown in SI table 1.

9

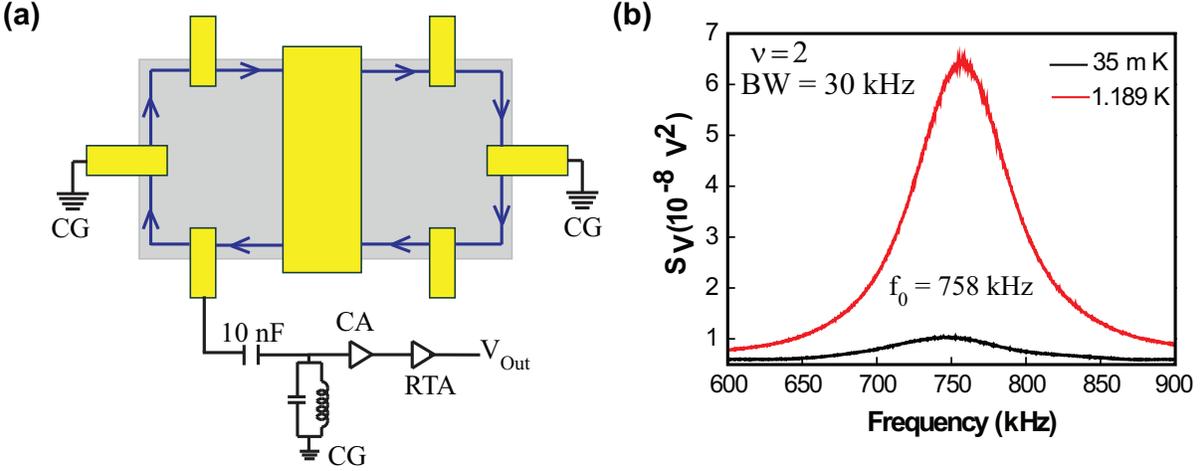

**SI Figure 5: Gain of amplification chain.** (**From temperature dependent thermal noise**) (**a**) Schematic of the measurement used to measure thermal noise. CA and RTA stand for "cold amplifier" and "room temperature amplifier", respectively. (**b**) The total noise measured by a spectrum analyzer for device-1 is plotted as a function of frequency at zero bias at temperature $\sim 35$ mK (black) and 1.189 K (red) ($RuO_2$ sensor reading) for filling $\nu = 2$. The resonance frequency is observed at $\sim 758$ kHz clearly visible in the plot. From peak value of these data, gain is calculated using $G = \sqrt{\frac{S_V(T_2) - S_V(T_1)}{4k_B(T_2 - T_1)BW}}$ (see SI section 2). Calculated gain was found $\sim 1425$, which is closely same as obtained from measurement of output voltage for a known input signal in quantum Hall states at resonance frequency.

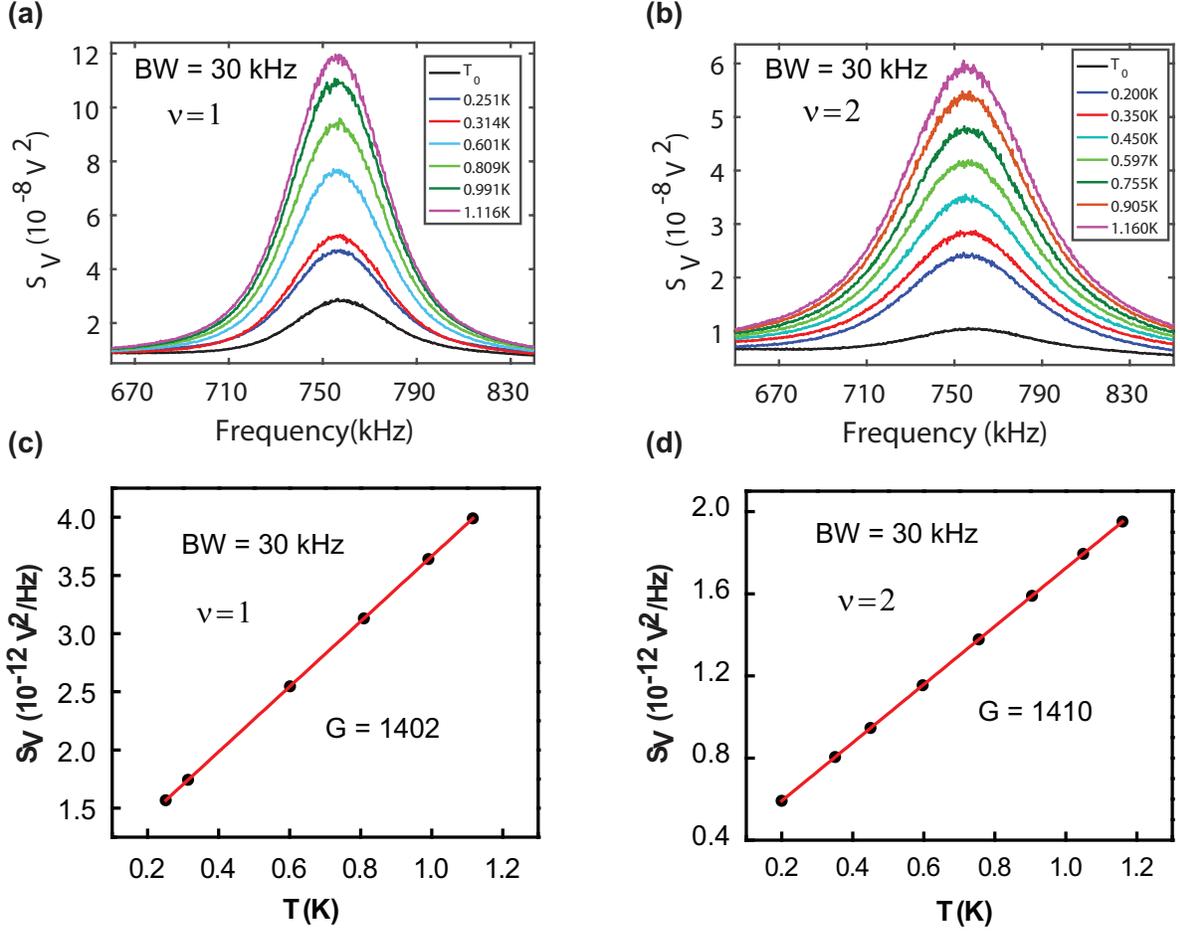

**SI Figure 6: Gain of amplification chain during measurement of device-3 (graphite back gated device)** (**a**) Noise measured by spectrum analyzer is plotted as a function of frequency at different temperature at $\nu = 1$ and similarly for (**b**) $\nu = 2$. (**c**) Symbols represent the plot of noise divided by bandwidth (BW) at resonance frequency as a function of temperature at $\nu = 1$. Solid red line is the linear fit of data and from the slope of this line, calculated gain was found to be equal to $\sim$1402. (**d**) Symbols represent the plot of noise divided by bandwidth (BW) at resonance frequency as a function of temperature at $\nu = 2$. Solid red line is the linear fit of data and from the slope of this line, calculated gain was found to be equal to $\sim$1410. From the intercept of these linear fit of $\nu = 1$ and 2, the $V_n$ and $i_n$ of amplifier were found to be $\sim 246 \frac{pV}{\sqrt{Hz}}$ and $\sim 24 \frac{fA}{\sqrt{Hz}}$, respectively. Thus, it can be seen that the error in gain is less than 1%.

| Filling Factor($\nu$) | $V_{IN}$ ($\mu$V) | $V_{OUT}$(mV) | Gain (G) = 2 ( $V_{OUT}$ / $V_{IN}$ ) | Gain from Jhonson Nyquist Noise |
|---|---|---|---|---|
| 2 | 39.97 | 28.56 | 1429.1 | |
| 6 | 39.03 | 27.80 | 1424.5 | 1425 |
| 10 | 37.54 | 26.84 | 1429.9 | |

SI Table 1: Gain of amplification chain.

# SI section-3: Electron temperature ($T_0$) determination

**Electron temperature for device-1 and device-2:** In this section we will discuss how we have extracted out the electron temperature of the devices. For device-1 and device-2 we have used coaxial line as input as shown in SI figure 2, where the $\sim 60$ ohm resistance and $\sim 1$ nF capacitance of the line work as low-pass filter. In this setup we have measured the thermal noise at different filling and from which we have extracted the electron temperature as shown in SI figure 8 and SI table 2. To cross check we have measured the shot noise at a graphene pn junction using same coaxial lines and extract out the electron temperature as shown in SI figure 9. The average electron temperature using coaxial line in our experiment was $T_0 \sim 40$ mK.

**Electron temperature for device-3(graphite back gate):** For device-3 (graphite back gated device) we have introduced extra low-pass filter made of 200 ohm and 1 nF capacitor in input line at the mixing chamber to lower the electron temperature which can be seen from the photograph of cold finger in SI figure 7. As we have mentioned earlier in SI section-2, that noise measured by spectrum analyzer at zero bias is given by

$$S_V = G^2(4k_B T R + V_n^2 + i_n^2 R^2)BW \tag{16}$$

If one plots the $\frac{S_V}{BW}$ as a function of temperature at fixed quantum Hall filling as plotted in SI figure 6c and 6d, the linear fit of the data will give an linear equation which relates the $\frac{S_V}{BW}$ to temperature T directly. Since the gain (from the slope) and intrinsic noise of amplification chain (from the intercept) is already known (see caption of SI figure 6), so now from the known value of measured noise at zero bias and base temperature, the corresponding electron temperature can be found directly using the following equation.

$$T_0 = \frac{\left( \left( \frac{S_V}{G^2 BW} \right) - (V_n^2 + i_n^2 R^2) \right)}{4k_B R} \tag{17}$$

As our measured value of noise at base temperature were $2.8180 \times 10^{-8} V^2$ at $\nu = 1$ which corresponds to $T_0 \sim 27$ mK. Similarily, at $\nu = 2$ measured noise was $1.0404 \times 10^{-8} V^2$, which also corresponds to $T_0 \sim 27$ mK.

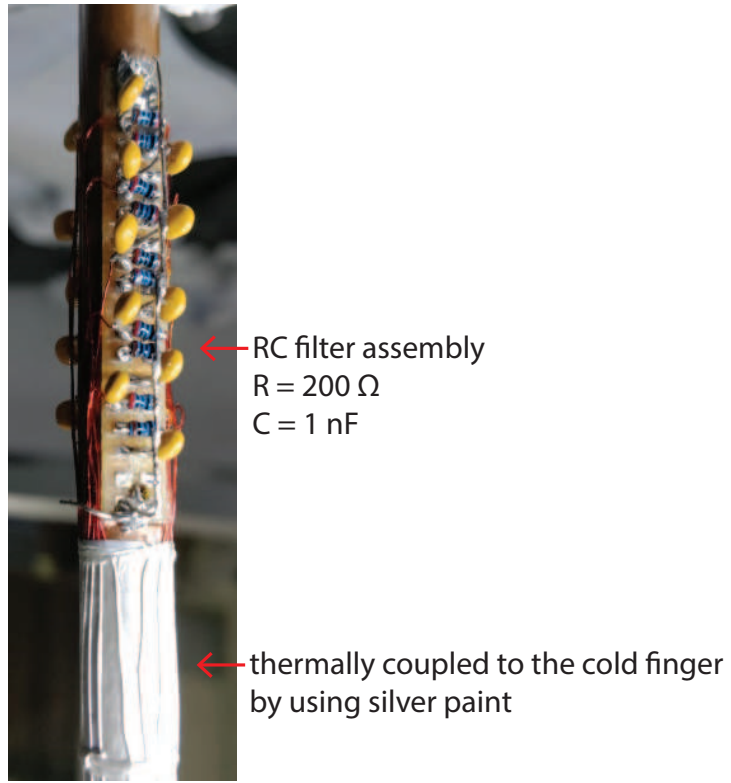

RC filter assembly
R = 200 Ω
C = 1 nF

thermally coupled to the cold finger
by using silver paint

SI Figure 7: RC filter assembly and thermal anchoring on the cold finger

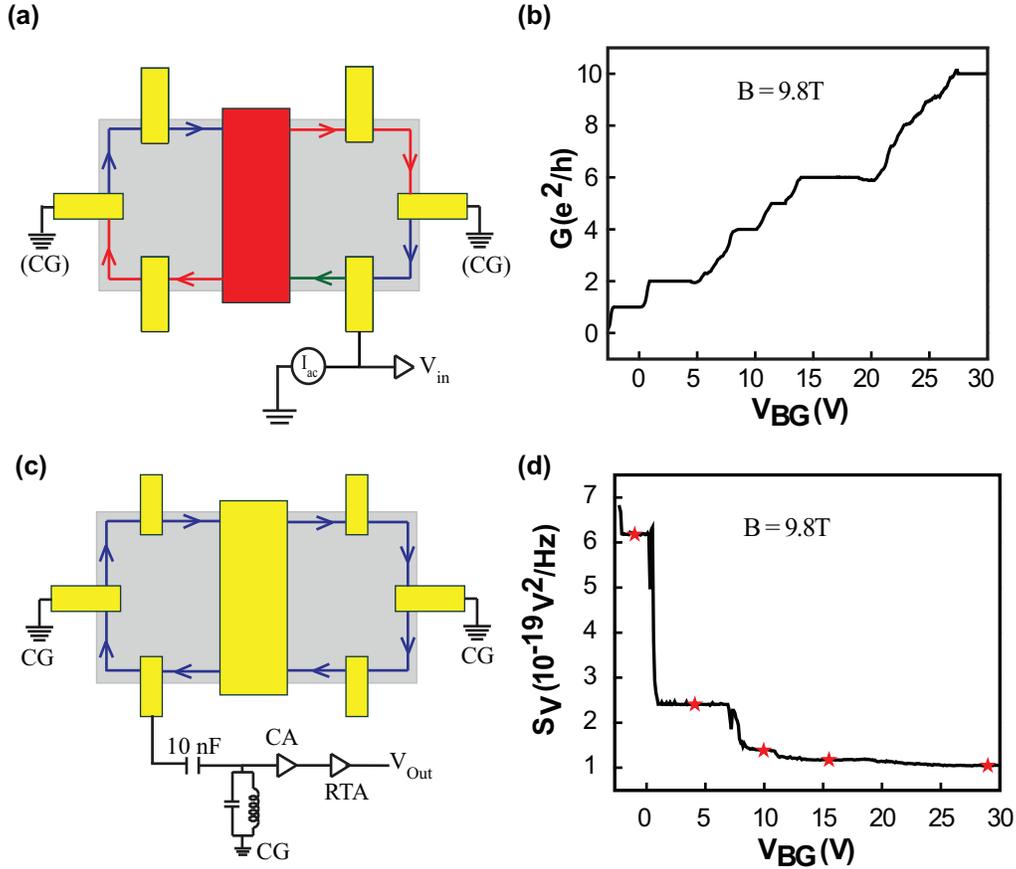

**SI Figure 8: Electron temperature ($T_0$) determination.** We have used two different methods to find the electron temperature $T_0$. (i) Measuring the thermal noise for different integer quantum Hall filling. (ii) Measuring the shot noise in p-n junction graphene device in integer quantum Hall regime.

(**From measuring the thermal noise for different integer quantum Hall filling:**) (**a**) Schematic of the measurement used to measure the conductance at low frequency (Lock-in at 228 Hz). (**b**) Gate response of Device-1 at B = 9.8T using low frequency Lock-in technique. Clear plateaus in conductance at 1, 2, 4, 5, 6, and 10 and less developed plateau at 3, 7, 8, and 9 are observed in unit of $e^2/h$. (**c**) Schematic of the measurement used to measure thermal noise. CA and RTA stand for "cold amplifier" and "room temperature amplifier", respectively. (**d**) The total noise measured by spectrum analyzer is plotted as a function of gate voltages at zero bias at base temperature of mixing chamber plate. Clear plateaus in these data are exactly at the same place as in conductance data. Red stars are the point, whose value we have taken to determine the electron temperature $T_0$. As mentioned in SI section-2, total noise measured by spectrum analyzer is $S_V = G^2(4k_BTR + V_n^2 + i_n^2R^2)BW$. Varying the filling factor will only change the R in this equation. So by measuring the total noise $S_V$ at three different fillings, we will get three linearly independent equation in three variables T, $V_n$, and $i_n$, which can be solved algebraically. Calculated values are shown in SI table 2.

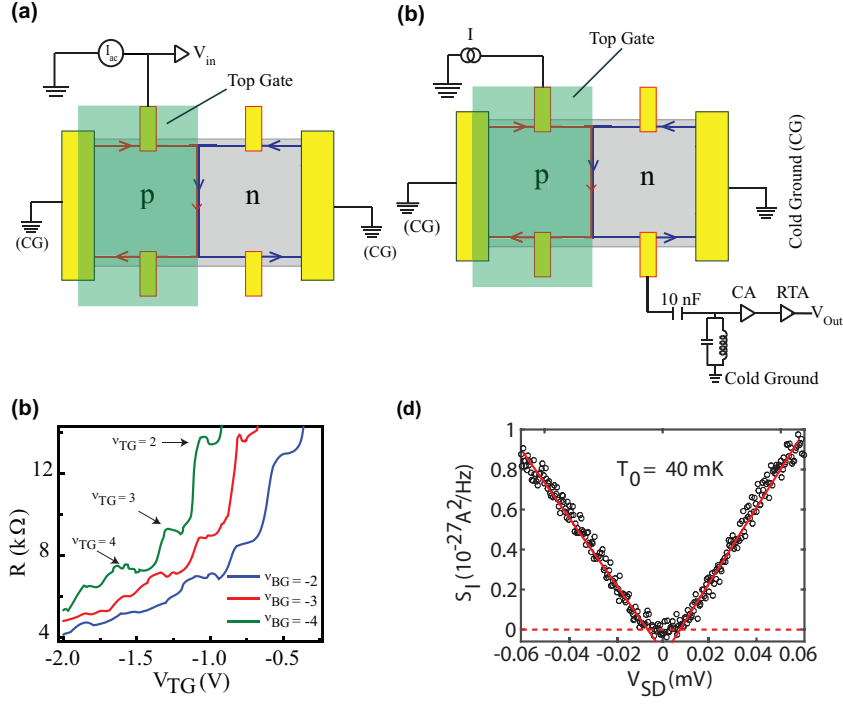

**SI Figure 9: Electron temperature ($T_0$) determination. (From shot noise measurement in a p-n junction of graphene device)** We have done the shot noise measurement in seperate p-n junction graphene device in integer quantum Hall regime using same measurement set-up which we have used for our device-1 and device-2. (**a**) Schematic of the measurement used to measure low frequency (228 Hz) conductance. (**b**) Schematic of the measurement used to measure the shot noise. CA and RTA stand for "cold amplifier" and "room temperature amplifier", respectively. Shot noise is measured using the spectrum analyzer. (**c**) Resistance $\left(\frac{V_{in}}{I_{ac}}\right)$ is plotted as a function of top gate voltage for the fixed filling of back gate of $\nu$ = -2 (blue), -3 (red), and -4 (green). For the given filling of back gate, we can see the plateaus as the top gate is scanned. At the plateau $\nu = (4, -4)$, we have measured the shot noise. (**d**) Symbols display the shot noise as a function of $V_{SD}$, where $V_{SD}$ is related to the applied DC current I as $I = G_{PNJ}V_{SD}$ with $G_{PNJ}$ as electrical conductance across p-n junction (7, 8). The horizonatl dashed line represents the zero noise line. Continuous red lines are the linear fit of these data at higher voltages whose linear extrapolation to zero noise gives $eV_{SD} = 2k_BT_0$ (9). In our case, the linear extrapolation to zero noise gives $V_{SD} \sim 6.95\mu V$. This corresponds to electron temperature $T_0 \sim 40$ mK, which is independent of gain of amplification chain.

| configuration of ν | | | $T_0$ (mK) | $V_n$ (nV/√Hz) | $I_n$ (fA/√Hz) |
|---|---|---|---|---|---|
| 1 | 2 | 4 | 44.7 | 0.3060 | 26.29 |
| 1 | 2 | 6 | 44.0 | 0.3066 | 26.31 |
| 1 | 2 | 10 | 44.3 | 0.3063 | 26.30 |
| 1 | 4 | 6 | 42.1 | 0.3072 | 26.37 |
| 1 | 4 | 10 | 43.7 | 0.3064 | 26.32 |
| 2 | 4 | 6 | 41.0 | 0.3075 | 26.50 |
| 2 | 4 | 10 | 43.4 | 0.3065 | 26.36 |
| 1 | 6 | 10 | 45.4 | 0.3061 | 26.26 |
| 2 | 6 | 10 | 45.9 | 0.3060 | 26.19 |

**SI Table 2: Electron temperature ($T_0$).** The average temperature $T_0$ calculated from above table is the $(43.8 \pm 1.4)$ mK. This corresponds to 3.2% error in the $T_0$.

# SI section-4: partition of current and contact resistance

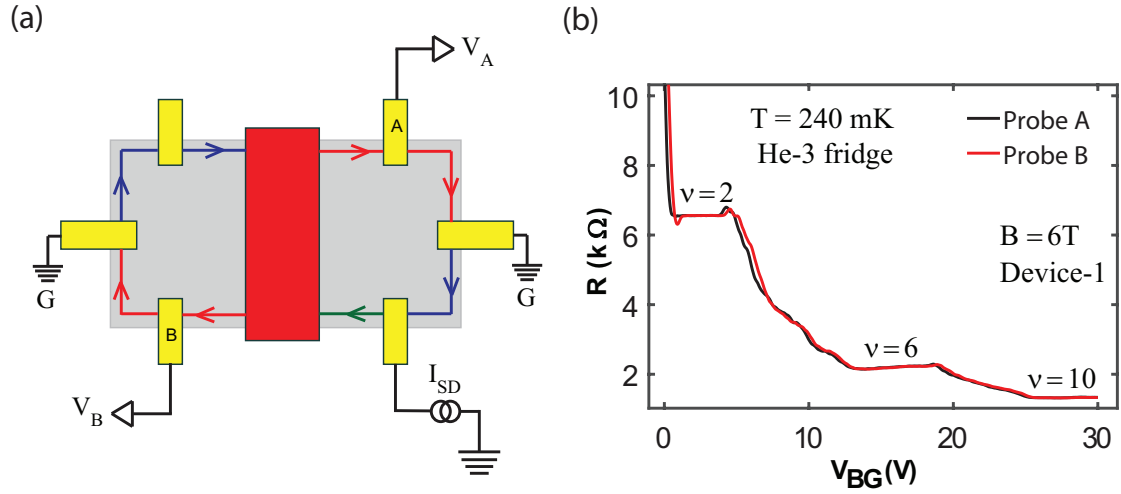

**SI Figure 10: Equi-partition of current in left and right moving chiral states:** For a given back gate voltage, filling in both arm of the device should be same. To confirm this we have measured the voltages in both arm of the device-1. (**a**) Current is applied through the source contact and the voltages are measured using probe 'A' and 'B', respectively. (**b**) Voltages divided by the applied current is plotted as a function of gate voltages measured at two different probes 'A' (black) and 'B' (red). It is clearly visible that the voltages measured at two probes are the same.

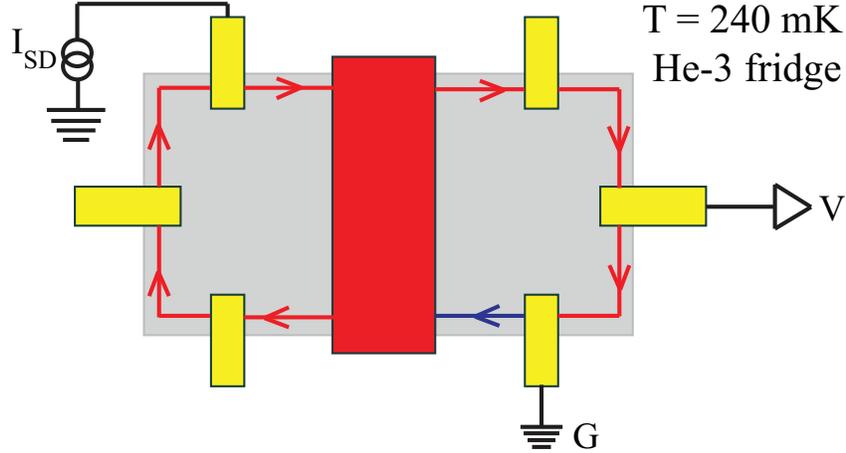

**SI Figure 11: Determination of contact resistance and source noise:** At finite DC bias, if the contacts are not transparent for the edge channels, it will give extra noise, known as source noise whose magnitude will be given as $2eI(1-t)$, where $t$ is the transmittance of source contact. To determine the contact resistance of source contact, we have used the three probe measurement technique. In our device configuration, if voltage probe $V$ is situated in path of hot edge for given chirality, it will measure a voltage $V_H = I.(R_0 + R_L + R_C)$ with $R_0$-quantum Hall resistance, $R_C$-contact resistance and $R_L$-line resistance of ground contact. On the other hand for opposite chirality, it will measure a voltage $V_C = I.R_L$. Differences of these two voltages will give us $I.(R_0 + R_C)$. So conductance is given by $\frac{1}{R_0+R_C}$ hence transmittance $t$ will be given by the ratio of actual resistance and measured resistance, i.e. $t = \frac{R_0}{R_0+R_C}$. Once the transmittance is known, one can easily determine the source noise $2eI(1-t)$. Since amplifier is situated in path of left moving edge channels, it will measure only part of the generated source noise. In particular, for our device configuration, amplifier will always see only the half of total no of edge channels leaving the floating reservoir. Hence it will always measure the only $\frac{1}{4}$th of the source noise. Values of the contact resistance, transmittance and maximum source noise is given in SI table 3 for device-1 and device-2 and in SI table 4 for device-3.

| Device | Filling Factor($\nu$) | Hot edge $V_H/I = (R_0 + R_L + R_C)(\Omega)$ | Cold edge $V_C/I = R_L$ ($\Omega$) | Contact Resistance $(V_H/I - V_C/I) - R_0$ in ($\Omega$) | Trans-mittance ( t ) | Source Noise / 4 ( $10^{-29} A^2$/ Hz) at $I_{max}$ |
|---|---|---|---|---|---|---|
| 1 | 1 | 26509 | 635 | 64 | 0.9975 | 0.0495 @ $I_{max}$ = 2.5 nA |
| | 2 | 13574 | 635 | 34 | 0.9974 | 0.0736 @ $I_{max}$ = 3.5 nA |
| | 6 | 4952 | 635 | 16 | 0.9963 | 0.2075 @ $I_{max}$ = 7.0 nA |
| 2 | 2 | 13570 | 624 | 41 | 0.9968 | 0.1536 @ $I_{max}$ = 6.0 nA |
| | 6 | 4939 | 624 | 13 | 0.9969 | 0.2232 @ $I_{max}$ = 6.0 nA |

**SI Table 3: Contact resistance and the Source noise.** Since the reflection was always less than 1% for each filling factor. So the measured source noise could be as maximum as about $0.2 \times 10^{-29} A^2 Hz^{-1}$ at maximum source current which is $\sim 80$ times smaller than the measured excess thermal noise. Make note that typical line resistance of He-3 fridge was 625-635 $\Omega$, which was measured also seperately.

| Device | Filling Factor($\nu$) | Measured resistance ($R_0 + R_L + R_C$) ($\Omega$) | Line resistance $R_L$ ($\Omega$) | Contact Resistance ($R_0 + R_L + R_C$) - $R_L$ - $R_0$ in ($\Omega$) | Trans-mittance ($t$) | Source Noise / 4 ($10^{-29} A^2$/ Hz) at $I_{max}$ |
|---|---|---|---|---|---|---|
| 3 | 1 | 26143 | 260 | 73 | 0.9972 | 0.0448 @ $I_{max}$ = 2.0 nA |
| | 4/3 | 19660 | 260 | 42 | 0.9978 | 0.0264 @ $I_{max}$ = 1.5 nA |
| | 2 | 13231 | 260 | 66 | 0.9949 | 0.0816 @ $I_{max}$ = 2.0 nA |

**SI Table 4: Contact resistance and the Source noise of device-3 (graphite back gate device).** Contact resistance of device-3 was measured in dilution refrigerator using low frequency Lock-in technique at 228 Hz. A constant current I was injected at source contact and voltage was measured at the same contact (See the SI figure 8(a)). This voltage probe will measure the voltage $I.(R_0 + R_L + R_C)$ with $R_0$-quantum Hall resistance, $R_C$-contact resistance and $R_L$-line resistance. So measured resistance will be $R_0 + R_L + R_C$. We have measured line resistance $R_L$ of 260 $\Omega$ seperately. After subtracting the line resistance, conductance will be equal to $\frac{1}{R_0 + R_C}$. Hence the transmittance $t$ will be given by $t = \frac{R_0}{R_0 + R_C}$. Once the transmittance is known, one can easily determine the source noise (see the caption of SI fig 11). So the maximum source noise at maximum current is much lower than the measured excess thermal noise (see SI figure 15).

# SI section-5: Dissipated power in the floating reservoir

As mentioned in the manuscript a DC current $I$, injected at the source contact (Fig. 1a of the manuscript), flows along the chiral edge towards the floating reservoir. The outgoing current from the floating reservoir splits into two equal parts, each propagating along the outgoing chiral edge from the floating reservoir to the cold grounds. The floating reservoir reaches a new equilibrium potential $V_M = \frac{I}{2\nu G_0}$ with the filling factor $\nu$ of graphene determined by the $V_{BG}$, whereas the potential of the source contact is $V_S = \frac{I}{\nu G_0}$. Thus, the power input to the floating reservoir is $P_{in} = \frac{1}{2}(IV_S) = \frac{I^2}{2\nu G_0}$, where the pre-factor of 1/2 results due to the fact that equal power dissipates at the source and the floating reservoirs in Fig. 1a of the manuscript. Similarly, the outgoing power from the floating reservoir is $P_{out} = \frac{1}{2}(2 \times \frac{I}{2}V_M) = \frac{I^2}{4\nu G_0}$. Thus, the resultant injected power dissipation in the floating reservoir due to joule heating is $J_Q = P_{in} - P_{out} = \frac{I^2}{4\nu G_0}$.

# SI section-6: Determination of temperature ($\mathbf{T_M}$) of floating reservoir

The resulting increase in the electronic temperature ($T_M - T_0$) of floating reservoir is determined from the excess thermal noise (*9–14*):

$$S_I = 2G^* k_B (T_M - T_0) \tag{18}$$

with

$$\frac{1}{G^*} = \frac{1}{G_L} + \frac{1}{G_R} \tag{19}$$

where $G_L$ and $G_R$ are the conductance of left and right channel respectively. So in our device structure,

$$\frac{1}{G^*} = \frac{1}{\nu G_0} + \frac{1}{\nu G_0} \tag{20}$$

hence

$$S_I = \nu k_B (T_M - T_0) G_0 \tag{21}$$

# SI section-7: Extended excess thermal noise data

In order to measure the excess thermal noise due to very small temperature increment, we have taken the data many times ($\sim 1000$) and we have shown the average data in the manuscript. In the SI figure 12a and 12b, we have shown one such raw data and the average one for $\nu = 1$ and 2, respectively.

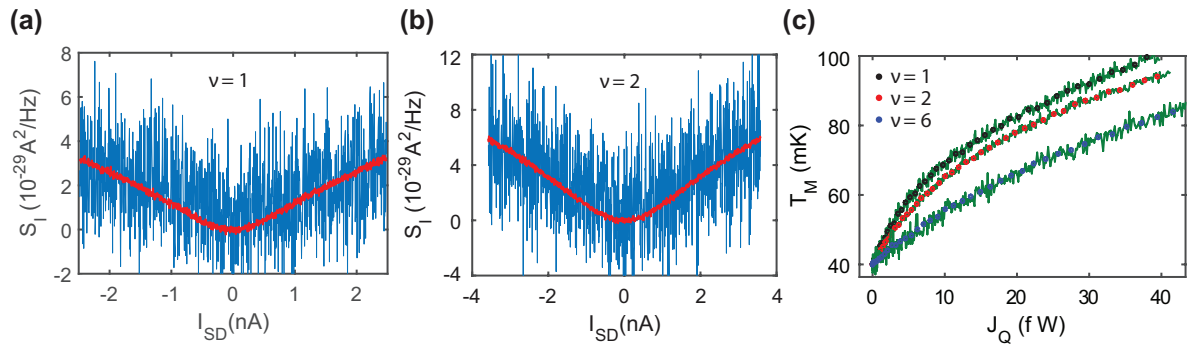

**SI Figure 12:** (**a,b**) Excess noise for single scan (blue) and the averaged (red) one for $\nu = 1$ and 2 at B = 9.8T respectively, for device-1. (**c**) Solid curves shows the data extracted directly from raw excess thermal noise data for device-1 at B = 9.8T for $\nu = 1, 2$, and 6 respectively. Solid symbols display the 9 point average of corresponding raw data which is shown in manuscript figure 2(d).

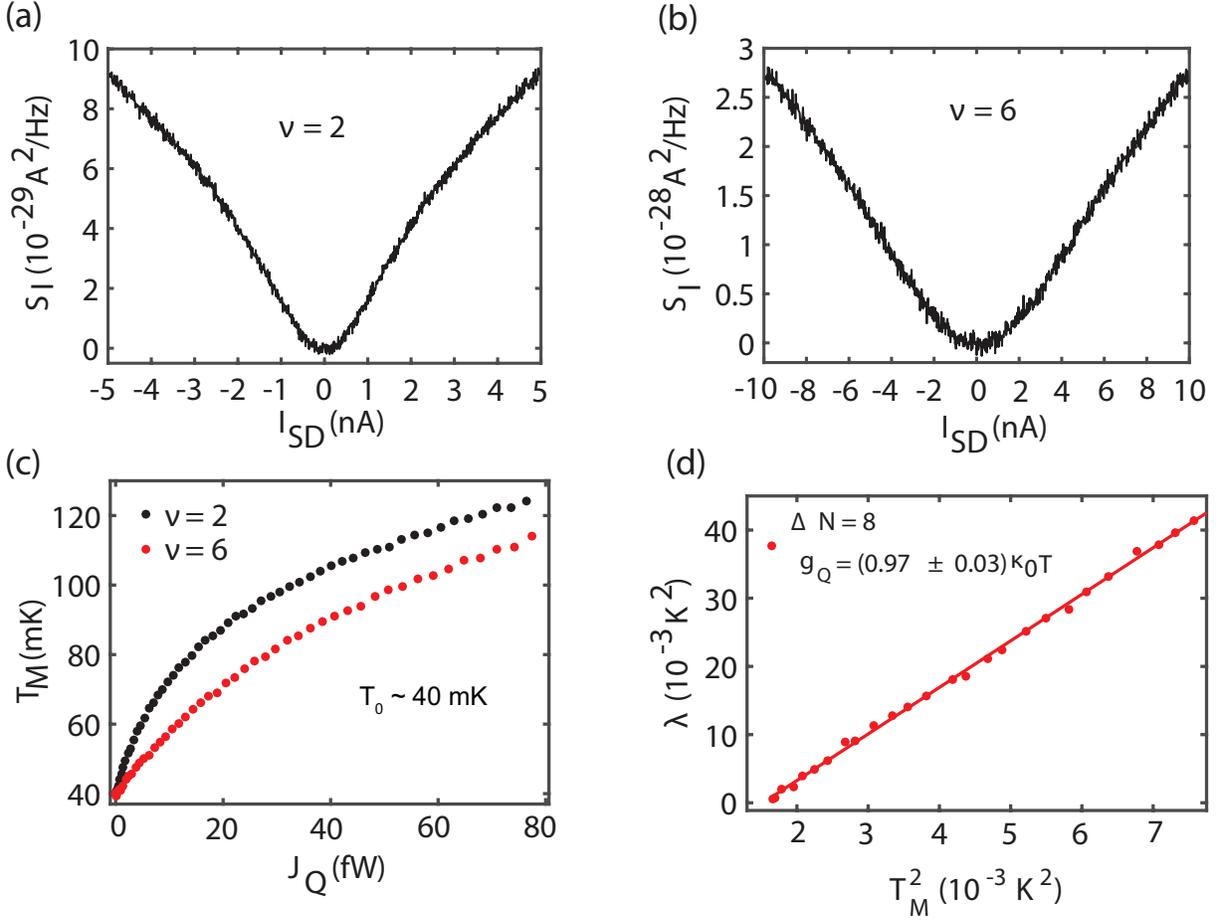

**SI Figure 13: Extended data of device-1 at B = 6T.** (**a**) Excess thermal noise $S_I$ is plotted as a function of source current $I_{SD}$ at filling $\nu = 2$. (**b**) Excess thermal noise $S_I$ is plotted as a function of source current $I_{SD}$ at filling $\nu = 6$. (**c**) Temperature, $T_M$ (it is extracted from the excess thermal noise shown in upper panel) of floating contact is plotted as a function of dissipated power $J_Q$ (obtained using $J_Q = \frac{I^2}{4\nu G_0}$) for two different filling factors $\nu = 2$ and 6, respectively. Symbols display the extracted temperature data using equation $S_I = \nu k_B(T_M - T_0)G_0$. (**d**) The $\lambda = \Delta J_Q/(0.5\kappa_0)$ is plotted as a function of $T_M^2$ for $\Delta N = 8$ (here $\Delta J_Q = J_Q(\nu_i, T_M) - J_Q(\nu_j, T_M)$). Solid circles display the data and the continuous red curve is the linear fit of these data. Slope of this linear fit is 7.76 which gives $g_Q = 0.97\kappa_0$T with specified error.

24

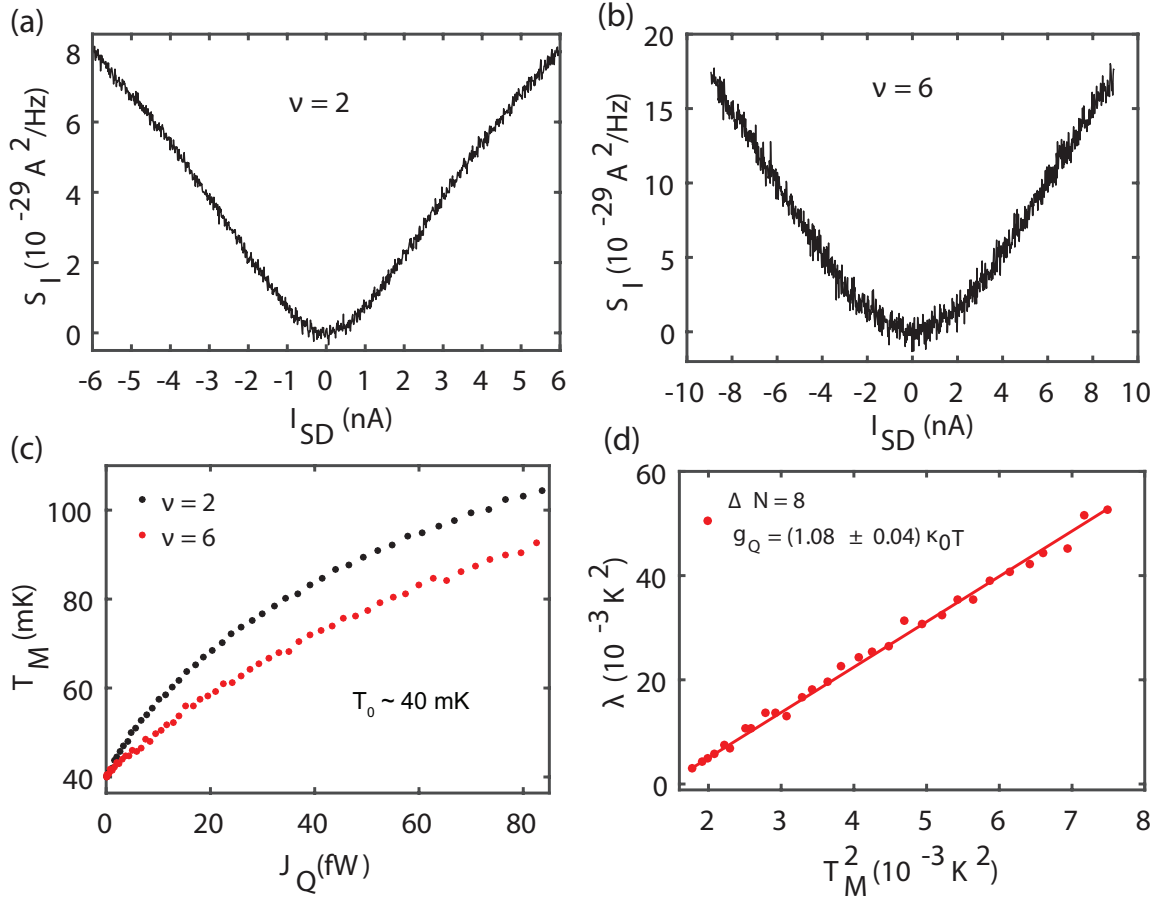

**SI Figure 14: Extended data of device-2 at B = 6T.** (**a**) Excess thermal noise $S_I$ is plotted as a function of source current $I_{SD}$ at filling $\nu = 2$. (**b**) Excess thermal noise $S_I$ is plotted as a function of source current $I_{SD}$ at filling $\nu = 6$. (**c**) Temperature $T_M$ (it is extracted from the excess thermal noise shown in upper pannel) of floating contact is plotted as a function of dissipated power $J_Q$ (obtained using $J_Q = \frac{I^2}{4\nu G_0}$) for two different filling factors $\nu = 2$ and 6, respectively. Symbols display the extracted temperature data using equation $S_I = \nu k_B (T_M - T_0) G_0$. (**d**) The $\lambda = \Delta J_Q/(0.5\kappa_0)$ is plotted as a function of $T_M^2$ for $\Delta N = 8$ (here $\Delta J_Q = J_Q(\nu_i, T_M) - J_Q(\nu_j, T_M)$). Solid circles display the data and the continuous red curve is the linear fit of these data. Slope of this linear fit is 8.64 which gives $g_Q = 1.08\kappa_0 T$ with specified error.

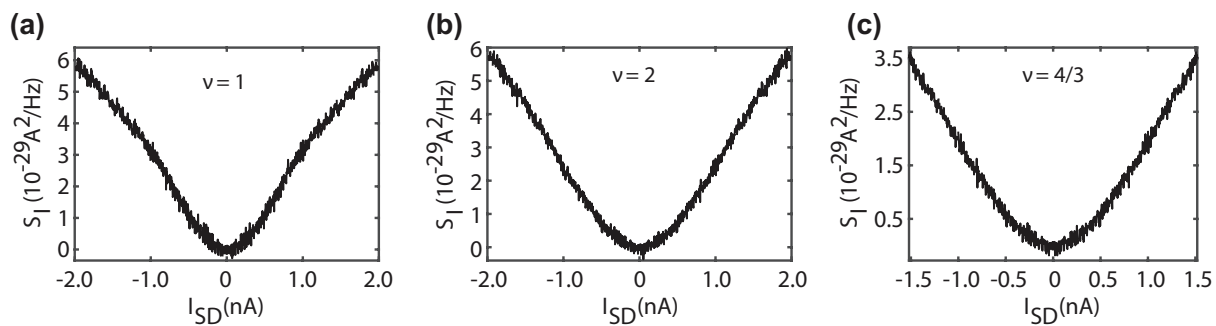

**SI Figure 15: Extended data of device-3 (graphite back gate) at B = 7T. (a)** Excess thermal noise $S_I$ is plotted as a function of source current $I_{SD}$ at filling $\nu = 1$. **(b)** Excess thermal noise $S_I$ is plotted as a function of source current $I_{SD}$ at filling $\nu = 2$ and similarly **(c)** Excess thermal noise $S_I$ is plotted as a function of source current $I_{SD}$ at filling $\nu = 4/3$.

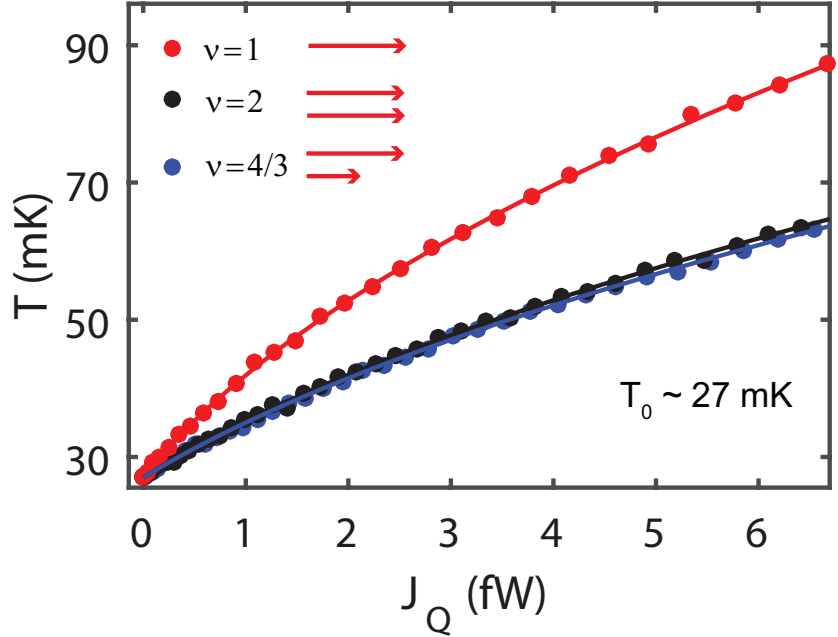

**SI Figure 16: Extended data of device-3 (graphite back gate) at B = 7T.** Temperature $T_M$ (extracted from excess thermal noise plotted in SI figure 15) of floating contact is plotted as a function of dissipated power $J_Q$ for filling factors $\nu = 1$, 2, and 4/3. Symbols display the extracted temperature data using equation $S_I = \nu k_B (T_M - T_0) G_0$. Here, the continuous lines are the best fit of data in accordance with equation $J_Q = 0.5N\alpha(T_M^2 - T_0^2)$ (here, we have not included the electron-phonon cooling). There are always two path from floating contact: left moving and right moving channels terminated at respective cold grounds. From fitting $\alpha$ were found to be 1.02, 2.02 and $2.08\kappa_0$ for $\nu = 1$, 2 and 4/3, respectively.

# SI section-8: Heat loss by Electron-phonon cooling

In this section, we will discuss about the heat transfer via electron-phonon cooling ($J_Q^{e-ph}$). In steady state, the total dissipated heat power in floating contact ($J_Q$) is equal to the sum of heat current carried by electronic channels ($J_Q^e$) and the heat transferred by hot electrons to the cold phonon bath ($J_Q^{e-ph}$) i.e. $J_Q = J_Q^e + J_Q^{e-ph}$, with $J_Q^e = 0.5N\alpha(T_M^2 - T_0^2)$ and $J_Q^{e-ph} = \beta(T_M^q - T_0^q)$. The temperature exponent of $5 - 5.8$ has been reported in GaAs 2DEG (*15*). However, the exponent of 4-6 is predicted depending on the nature of disorders in the floating contact as well as the dimensionality of the phonons involved (*16,17*). In our devices, we found the temperature exponent of $\sim 4$ and $\sim 6$ for $SiO_2$ and graphite back gate, respectively as shown in SI figure 17.

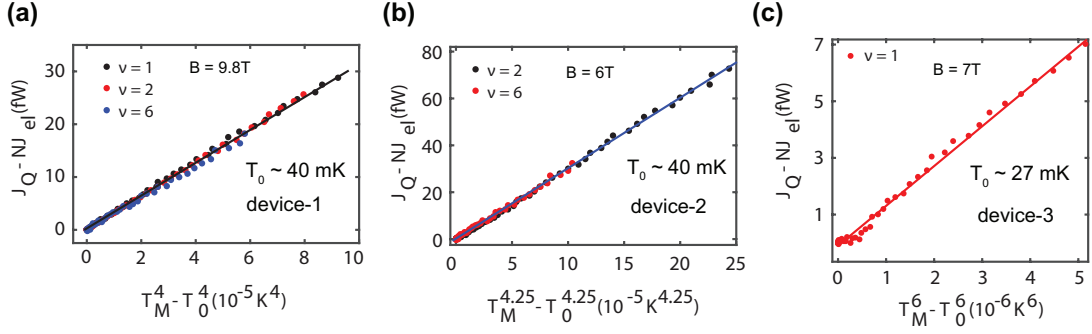

**SI Figure 17: Heat loss by electron-phonon coupling.** (**a**) Solid circles display the electron-phonon contribution of the heat loss, $J_Q^{e-ph} = J_Q - J_Q^e$ with $J_Q^e = N J_{el} = N 0.5 \kappa_0 (T_M^2 - T_0^2)$ electronic heat flow through $N$ chiral edge channels, as a function of $(T_M^4 - T_0^4)$ for $\nu = 1, 2$, and 6 for device-1. The solid line is the linear fit with a slope, $\beta \sim 0.31 nW/K^4$. (**b**) Solid circles display the electron-phonon contribution of the heat loss as a function of $(T_M^{4.25} - T_0^{4.25})$ for $\nu = 1$, and 2 for device-2. The solid line is the linear fit with a slope, $\beta \sim 0.30 nW/K^{4.25}$. (**c**) Here, soild circles display the same for device-3 (graphite back gated device) for $\nu = 1$, but it increases with $(T_M^6 - T_0^6)$ with $\beta \sim 1.54 nW/K^6$. Here, we have not plotted the heat loss to phonon for $\nu = 2$ and $4/3$, because at these filling electron-phonon parts were very small. In this device (graphite back gate) the floating reservoir is isolated from the graphite back gate by the bottom hBN but in the case of $SiO_2/Si$ back gated device the floating reservoir is in touch with the $SiO_2/Si$ substrate as the bottom hBN is completely etched out, which might be the reason for different temperature exponent.

# SI section-9: Accuracy of the thermal conductance measurement

In this section, we will discuss about the accuracy of measured thermal conductance, which requires the careful determination of the parameters of the system. For example: (i) the exact value of total gain of the amplification chain; (ii) the precise measurement of the electron temperature ($T_0$) and (iii) estimation of generated shot noise due to the finite contact resistance at the graphene QH and electrode junctions. The total voltage gain of our amplification chain is $\sim$1402-1425 measured with known input signal as well as measuring the temperature dependent thermal noise in QH plateaus (SI), where the

inaccuracy of the gain $\sim 1\%$. The electron temperature, $T_0 \sim 40$mK for device-1 and device-2, was measured in two ways; measuring the thermal noise (together with voltage and current noise of the amplifier) at different filling factors and measuring the shot noise in a graphene pn junction in QH regime. The inaccuracy in our electron temperature measurement is less than 4% as can be seen in SI table 2. Since the inaccuracy in gain is $\sim 1\%$, so it will not affect measured thermal conductance very much. On the other hand, inaccuracy in electron temperature is $\sim 4\%$, which can be source of major inaccuracy in extracted thermal conductance. So we have calculated the thermal conductance at $T_0 = 40$ mK and 45 mK. The averaged value was found $(1.00 \pm 0.05)\kappa_0 T$ at $T_0 = 40$ mK and $(0.94 \pm 0.04)\kappa_0 T$ at $T_0 = 45$ mK. So the change in the electron temperature of 12.5% corresponds to the 6% change in thermal conductance as shown in SI table 5. Since the inaccuracy in electron temperature is less than 4%, inaccuracy in extracted thermal conductance will be even much lesser than 6%. The estimated shot noise due to the finite contact resistance in our device is at least one order of magnitude smaller compared to measured excess thermal noise as shown in SI table 3 and table 4. The heated up electrons must dwell in the floating reservoir for a time longer than the electron-electron energy relaxation time in order to achieve a quasi-equilibrium temperature $T_M$, which can be easily justified in our devices. Since the typical electron-electron interaction at low temperature is $\sim 10$ ns in gold like material (*18*), which is very much lower than typical dwell time ($t_{dwell}$) $\sim 20$ $\mu s$ found in our devices. The $t_{dwell}$ inside micron-size floating contact is evaluated from the equation $t_{dwell} = \frac{D_E \Omega h}{N}$ (*10, 19*), with $D_E$ the electronic density of states per unit volume per unit energy and $\Omega$ the volume of micron-size floating contact and N the number of channel leaving the floating contact. In our devices, for the typical volume of floating contact of $\approx 5$ $\mu m^3$ and typical density of states for gold $D_E \simeq 1.14 \times 10^{47} J^{-1} m^{-3}$, we found $t_{dwell} \simeq \frac{40 \mu s}{N}$.

| Device | Magnetic field B ( T ) | Configuration | Thermal Conductance $g_Q$ ( in unit of $\kappa_0 T$) | |
|---|---|---|---|---|
| | | | $T_0 = 40$ mK | $T_0 = 45$ mK |
| 1 | B = 9.8 T | $\nu = 1$ and 2 | 0.96 | 0.90 |
| | | $\nu = 2$ and 6 | 0.99 | 0.92 |
| | B = 6.0 T | $\nu = 2$ and 6 | 0.97 | 0.91 |
| 2 | B = 6.0 T | $\nu = 2$ and 6 | 1.08 | 1.01 |

SI Table 5: **Change in thermal conductance for different electron temperature** $T_0$**.** The average value of electronic thermal conductance calculated from above table is the $(1.00 \pm 0.05)\kappa_0 T$ at $T_0 = 40$ mK and $(0.94 \pm 0.04)\kappa_0 T$ at $T_0 = 45$ mK. It is clear from above table that the change in the temperature of 12.5% corresponds to the 6% change in thermal conductance.



\end{document}